\documentclass[showpacs,amsfonts,aps,prc,floatfix,eqsecnum,groupedaddress,]%
{revtex4}
\usepackage{amsmath,amssymb,amsthm}
\usepackage{graphicx}
\usepackage{bm}
\usepackage{color}

\DeclareMathAlphabet{\mathpzc}{OT1}{pzc}{m}{it}

\begin{document}

\renewcommand{\textfraction}{0.00}

% Useful macros:

\newcommand{\vAi}{{\cal A}_{i_1\cdots i_n}} 
\newcommand{\vAim}{{\cal A}_{i_1\cdots i_{n-1}}} 
\newcommand{\vAbi}{\bar{\cal A}^{i_1\cdots i_n}}
\newcommand{\vAbim}{\bar{\cal A}^{i_1\cdots i_{n-1}}}
\newcommand{\htS}{\hat{S}} 
\newcommand{\htR}{\hat{R}}
\newcommand{\htB}{\hat{B}} 
\newcommand{\htD}{\hat{D}}
\newcommand{\htV}{\hat{V}} 
\newcommand{\cT}{{\cal T}} 
\newcommand{\cM}{{\cal M}} 
\newcommand{\cMs}{{\cal M}^*}
\newcommand{\vk}{\vec{\mathbf{k}}}
\newcommand{\bk}{\bm{k}}
\newcommand{\kt}{\bm{k}_\perp}
\newcommand{\kp}{k_\perp}
\newcommand{\km}{k_\mathrm{max}}
\newcommand{\vl}{\vec{\mathbf{l}}}
\newcommand{\bl}{\bm{l}}
\newcommand{\bK}{\bm{K}} 
\newcommand{\bb}{\bm{b}} 
\newcommand{\qm}{q_\mathrm{max}}
\newcommand{\vp}{\vec{\mathbf{p}}}
\newcommand{\bp}{\bm{p}} 
\newcommand{\vq}{\vec{\mathbf{q}}}
\newcommand{\bq}{\bm{q}} 
\newcommand{\qt}{\bm{q}_\perp}
\newcommand{\qp}{q_\perp}
\newcommand{\bQ}{\bm{Q}}
\newcommand{\vx}{\vec{\mathbf{x}}}
\newcommand{\bx}{\bm{x}}
\newcommand{\tr}{{{\rm Tr\,}}} 
\newcommand{\bc}{\textcolor{blue}}
\newcommand{\rc}{\textcolor{red}}

\newcommand{\beq}{\begin{equation}}
\newcommand{\eeq}[1]{\label{#1} \end{equation}} 
\newcommand{\ee}{\end{equation}}
\newcommand{\bea}{\begin{eqnarray}} 
\newcommand{\eea}{\end{eqnarray}}
\newcommand{\beqar}{\begin{eqnarray}} 
\newcommand{\eeqar}[1]{\label{#1}\end{eqnarray}}
 
\newcommand{\half}{{\textstyle\frac{1}{2}}} 
\newcommand{\ben}{\begin{enumerate}} 
\newcommand{\een}{\end{enumerate}}
\newcommand{\bit}{\begin{itemize}} 
\newcommand{\eit}{\end{itemize}}
\newcommand{\ec}{\end{center}}
\newcommand{\bra}[1]{\langle {#1}|}
\newcommand{\ket}[1]{|{#1}\rangle}
\newcommand{\norm}[2]{\langle{#1}|{#2}\rangle}
\newcommand{\brac}[3]{\langle{#1}|{#2}|{#3}\rangle} 
\newcommand{\hilb}{{\cal H}} 
\newcommand{\pleft}{\stackrel{\leftarrow}{\partial}}
\newcommand{\pright}{\stackrel{\rightarrow}{\partial}}

%%%%%%%%%%%%%%%%%%%%%%%%% Front Matter %%%%%%%%%%%%%%%%%%%%%%%%%%%%%%
\title{Magnetic and electric contributions to the energy loss in a dynamical 
QCD medium}
\author{Magdalena Djordjevic}
\email[Correspond to\ ]{magda@ipb.ac.rs}
\affiliation{Institute of Physics Belgrade, University of Belgrade, Serbia}

\begin{abstract}
The computation of radiative energy loss in a finite size QCD medium with 
dynamical constituents is a key ingredient for obtaining reliable predictions 
for jet quenching in ultra-relativistic heavy ion collisions. It was 
previously shown that energy loss in dynamical QCD medium is significantly 
higher compared to static QCD medium. To understand this difference, we here 
analyze magnetic and electric contributions to energy loss in dynamical QCD 
medium. We find that the significantly higher energy loss in the dynamical 
case is entirely due to appearance of magnetic contribution in the dynamical 
medium. While for asymptotically high energies, the energy loss in static and 
dynamical medium approach the same value, we find that the 
physical origin of the energy loss in these two cases is different. 
\end{abstract}

\date{\today} 
 

\maketitle
%%%%%%%%%%%%%%%%%%%%%%%%%%%%%%%%%%%%%%%%%%%%%%%%%%%%%%%%%%%%%%%%%%%%%%%

%%%%%%%%%%%%%%%%%%%%%%%%%%%%%%%%%%%%%%%%%%%%%%%%%%%%%%%%%%%%%%%%%%%%%%%
\section{Introduction}
\label{sec1}
%%%%%%%%%%%%%%%%%%%%%%%%%%%%%%%%%%%%%%%%%%%%%%%%%%%%%%%%%%%%%%%%%%%%%%%

Jet suppression~\cite{Bjorken} is considered to be a powerful tool to study 
the properties of a QCD medium created in ultra-relativistic heavy ion 
collisions~\cite{Brambilla,Gyulassy,DBLecture} . The suppression results from 
the energy loss of high energy partons moving through the 
plasma~\cite{suppression,BDMS,BSZ,KW:2004}. Therefore, reliable computations 
of jet energy loss mechanisms are essential for the reliable predictions of 
jet suppression. In~\cite{MD_PRC,DH_PRL}, we developed a theoretical formalism 
for the calculation of the first order in opacity light and heavy quark 
radiative energy loss in dynamical QCD medium (see also a
viewpoint~\cite{Gyulassy_viewpoint}). That study, which incorporates dynamical 
effects in realistic finite size QCD medium, enables us to provide the most 
reliable computations of the energy loss in QGP so far. This work has shown 
that the energy loss in dynamical medium is significantly larger (50-70\%), 
compared to the energy loss in static QCD medium.

The goal of this paper is to determine what is the origin of the observed 
significant energy loss increase in the case of dynamical QCD medium. To that 
end, we here analyze magnetic and electric contributions to the energy loss in 
the dynamical QCD medium, under conditions relevant at RHIC and LHC experiments. 
We furthermore analyze what happens for asymptotically large energies, where it 
was previously shown that the energy loss in static and dynamical QCD medium 
takes the same value~\cite{MD_PRC,DH_PRL}. 

%%%%%%%%%%%%%%%%%%%%%%%%%%%%%%%%%%%%%%%%%%%%%%%%%%%%%%%%%%%%%%%%%%%%%%%%
\section{Contributions to radiative energy loss in a dynamical QCD medium} 
\label{sec2}
%%%%%%%%%%%%%%%%%%%%%%%%%%%%%%%%%%%%%%%%%%%%%%%%%%%%%%%%%%%%%%%%%%%%%%%%

We here study the importance of electric and magnetic 
contributions to the medium induced radiative energy loss in a finite size 
dynamical QCD medium. In~\cite{MD_PRC,DH_PRL} it was shown that the radiative 
energy loss in a finite size dynamical QCD medium is given by: 
%
%\begin{widetext}
\beqar
\frac{\Delta E^{\mathrm{dyn}}}{E} 
= 2 \frac{C_R \alpha_s}{\pi}\,\frac{L}{\lambda_\mathrm{dyn}}  
    \int dx \,\frac{d^2k}{\pi} \,\frac{d^2q}{\pi} \, v(\bq)
        \left(1-\frac{\sin{\frac{(\bk{+}\bq)^2+\chi}{x E^+} \, L}} 
    {\frac{(\bk{+}\bq)^2+\chi}{x E^+}\, L}\right) %\nonumber \\
    %&& \; \; \; \times \,
    \frac{(\bk{+}\bq)}{(\bk{+}\bq)^2+\chi}
    \left(\frac{(\bk{+}\bq)}{(\bk{+}\bq)^2+\chi}
    - \frac{\bk}{\bk^2+\chi}
    \right) ,
\eeqar{DeltaEDyn}
%\end{widetext}
%
Here $L$ is the length of the finite size dynamical QCD medium and $E$ is the 
jet energy. $\bk$ is transverse momentum of radiated gluon, while $\bq$ is 
transverse momentum of the exchanged (virtual) gluon.  
$\alpha_s = \frac{g^2}{4 \pi}$ is coupling constant and $C_R=\frac{4}{3}$. 
$v(\bq)$ is the effective crossection in dynamical QCD medium and 
$\lambda_\mathrm{dyn}^{-1} \equiv C_2(G) \alpha_s T = 3 \alpha_s T$ ($C_2(G)=3$) 
is defined as ``dynamical mean free path'' (see also~\cite{DH_Inf}). 
$\chi\equiv M^2 x^2 + m_g^2$, where $x$ is the longitudinal 
momentum fraction of the heavy quark carried away by the emitted gluon, $M$ is
the mass of the heavy quark, $m_g=\mu_E/\sqrt 2$ is the effective mass for 
gluons with hard momenta $k\gtrsim T$ and $\mu_E$ is the Debye mass. We 
assume constant coupling $g$.

The effective crossection $v(\bq)$ can be written in the following form
\beqar
v(\bq) &=& v_L(\bq) + v_T(\bq) \, ,
\eeqar{vLT}
where $v_L(\bq)$ ($v_T(\bq)$) is longitudinal (transverse) contribution to the 
effective crossection, given by~\cite{Aurenche,MD_PRC}
\beqar
v_{L}(\bq) &=& \frac{1}{\bq^2+{\rm Re}\Pi_{L}(\infty)} - 
\frac{1}{\bq^2+{\rm Re}\Pi_{L}(0)} \nonumber \\
v_{T}(\bq) &=& \frac{1}{\bq^2+{\rm Re}\Pi_{T}(0)} - 
\frac{1}{\bq^2+{\rm Re}\Pi_{T, \,L}(\infty)} \, .
\eeqar{vTL}
$\Pi_L$ and $\Pi_T$ are HTL gluon self energies, given by\cite{Heinz_AP}
\bea
\Pi_L (l) = \mu_E^2 \left[ 1 - y^2 - \frac{y (1{-}y^2)}{2} 
\ln\left(\frac{y{+}1}{y{-}1}\right)\right],
\qquad
%\\
\label{PiT}
\Pi_T (l) &=& \mu_E^2 \left[ \frac{y^2}{2} + \frac{y (1{-}y^2)}{4} 
\ln\left(\frac{y{+}1}{y{-}1}\right)\right],
%\label{PiL}
\eea
where $y \equiv l_0/|\vl|$ ($l$ is the momentum of the gluon) and $\mu_E$ is 
Debye screening mass.

By using the following properties of the HTL self-energy of the
gluon \cite{Pisarski}
\beqar
&&{\rm Re}\,\Pi_{T,\,L}(\infty) =\mu^2_\infty \nonumber\\
&&{\rm Re}\,\Pi_{T}(0) =0\nonumber\\
&&{\rm Re}\,\Pi_{L}(0) =\mu_E^2 ,
\eeqar{PiValues}
where  $\mu_\infty\equiv \mu_E/\sqrt 3$ is the gluon thermal mass, we obtain
\beqar
v_L(\bq) &=& \frac{1}{(\bq^2+\mu_\infty^2)} -  \frac{1}{(\bq^2+\mu_E^2)} = 
\frac{\mu_E^2-\mu_\infty^2}{(\bq^2+\mu_\infty^2) (\bq^2+\mu_E^2)} \, , \nonumber\\
v_T(\bq) &=& \frac{1}{\bq^2} - \frac{1}{(\bq^2+\mu_\infty^2)} = 
\frac{\mu_\infty^2}{\bq^2 (\bq^2+\mu_\infty^2)} \,.
\eeqar{vTL_fin}

After replacing the effective crossection in Eq.~(\ref{DeltaEDyn}), with 
expressions from Eqs.~(\ref{vTL_fin}), we obtain magnetic (transverse) and 
electric (longitudinal) contributions to the energy loss:
\beqar
\frac{\Delta E^{dyn}_{\mathrm{M,\, E}}}{E} 
&=& \frac{C_R \alpha_s}{\pi}\,\frac{L}{\lambda_\mathrm{dyn}}  
    \int dx \,\frac{d^2k}{\pi} \,\frac{d^2q}{\pi} \, v_{T, \, L}(\bq)\,
    \left(1-\frac{\sin{\frac{(\bk{+}\bq)^2+\chi}{x E^+} \, L}} 
    {\frac{(\bk{+}\bq)^2+\chi}{x E^+}\, L}\right) \nonumber \\
    && \; \; \; \times \,
    2 \, \frac{(\bk{+}\bq)}{(\bk{+}\bq)^2+\chi}
    \left(\frac{(\bk{+}\bq)}{(\bk{+}\bq)^2+\chi}
    - \frac{\bk}{\bk^2+\chi}
    \right) ,
\eeqar{DeltaEMagnElec}

In the above equation $ v_{T}(\bq)$ under the integral corresponds to the 
magnetic contribution ($\Delta E^{dyn}_{\mathrm{M}}$), while $ v_{L}(\bq)$ 
corresponds to the electric contribution ($\Delta E^{dyn}_{\mathrm{E}}$). From 
Eq.~(\ref{vTL_fin}), we see that Debye screening ($\mu_E$) renders the 
longitudinal gluon exchange ($v_L(\bq)$) infrared finite, leading to a finite 
electric contribution to the energy loss Eq.~(\ref{DeltaEMagnElec}). The 
transverse gluon exchange ($v_T(\bq)$ in Eq.~(\ref{vTL_fin})) causes a 
well-known logarithmic singularity \cite{Le_Bellac}, due to the absence of a 
magnetic screening. However, one should note that the magnetic contribution 
($\Delta E^{dyn}_{\mathrm{M}}$) has to be finite as well, since 
in~\cite{DH_Inf,DH_PRL,MD_PRC} it was shown that the total energy loss in 
dynamical QCD medium is finite. Therefore, we can use Eqs.~(\ref{vTL_fin}) 
and~(\ref{DeltaEMagnElec}) to numerically study electric and magnetic 
contributions to dynamical energy loss, which will be done in 
Section~\ref{sec4}. 

%%%%%%%%%%%%%%%%%%%%%%%%%%%%%%%%%%%%%%%%%%%%%%%%%%%%%%%%%%%%%%%%%%%%%%%%
\section{Asymptotic limit of high energies} 
\label{sec3}
%%%%%%%%%%%%%%%%%%%%%%%%%%%%%%%%%%%%%%%%%%%%%%%%%%%%%%%%%%%%%%%%%%%%%%%%

We here analyze magnetic and electric contributions to the dynamical energy 
loss in the limit of high energies. The limit is obtained from 
Eq.~(\ref{DeltaEMagnElec}) when $E^+ \approx 2 E \rightarrow \infty$. In 
such a limit finite mass effects are negligible. Additionally, 
$\km= 2 E \sqrt{x (1-x)} \rightarrow \infty$ as well, which enables us to 
introduce a substitution $\bk^\prime \equiv \bk + \bq$ in 
Eq.~(\ref{DeltaEMagnElec}). With these simplifications, 
Eq.~(\ref{DeltaEMagnElec}) becomes
\beqar
\frac{\Delta E^{\mathrm{dyn}}_{\mathrm{M,\, E}}}{E} 
&=& \frac{C_R \alpha_s}{\pi}\,\frac{L}{\lambda_\mathrm{dyn}}  
    \int dx \,\frac{d^2k^\prime}{\pi} \,\frac{d^2q}{\pi} \, 
    v_{T, \, L}(\bq)\,
\, \frac{ 2 \, \bq \cdot (\bq-\bk^\prime)}
{ \bk^{\prime 2} \, (\bk^\prime-\bq)^2}\, 
\left(1- \frac{\sin (\frac{\bk^{\prime 2} \, L}{2xE})}
{\frac{\bk^{\prime 2} \, L}{xE^+}} \right) \,,
\eeqar{DeltaEDyn0}
where  $v_{T, \, L}(\bq)$ is given by Eq.~(\ref{vTL_fin}).

After performing the angular integration, taking the derivative over 
distance $L$ of the above expression and performing the integral over 
$\bk^{\prime 2}$, we obtain
\beqar
\frac{1}{E} \frac{dE^{\mathrm{dyn}}_{\mathrm{M,\, E}}}{dL}
&=& 2 \, \frac{C_R \alpha_s}{\pi}\,
\frac{1}{\lambda_\mathrm{dyn}}  
   \int_0^{q_{max}^2} d \bq^2 \,  \int_0^1 dx \, 
v_{T, \, L}(\bq)\,
\left(\gamma - {\rm Ci}\left(\frac{L \bq^2}{2 E x}\right) + 
\ln \left( \frac{L \bq^2}{2 E x} \right) \right)
\, ,
\eeqar{dEdl_xq} 
where $\gamma \approx 0.577216$ is Euler's constant and ${\rm Ci} (y)$ is 
the cosine integral function. 

Finally after performing the integration over $x$, we obtain (in the limit when 
$E \rightarrow \infty$)
\beqar
\frac{1}{E} \frac{dE^{\mathrm{dyn}}_{\mathrm{M,\, E}}}{dL} &=&
\frac{C_R \alpha_s}{2\, E}\,
\frac{L}{\lambda_\mathrm{dyn}} 
    \int_0^{q_{max}^2} d \bq^2 \, \bq^2 \, v_{T, \, L}(\bq) \, .
\eeqar{dEdl_LPM} 

Therefore, for the asymptotically large jet energies, the Eq.~(\ref{dEdl_LPM}) 
reduces to
\beqar
\frac{1}{E} \frac{dE^{\mathrm{dyn}}_{\mathrm{M}}}{dL} &=&
\frac{C_R \alpha_s}{2\, E}\, \frac{L}{\lambda_\mathrm{dyn}} \, \mu_\infty^2 \,  
    \ln \left( \frac{4 E T}{\mu_\infty^2} \right) 
\, , \nonumber \\
\frac{1}{E} \frac{dE^{\mathrm{dyn}}_{\mathrm{E}}}{dL} &=&
\frac{C_R \alpha_s}{2\, E}\, \frac{L}{\lambda_\mathrm{dyn}} \, 
\left[ \mu_E^2 \, \ln \left( \frac{4 E T}{\mu_E^2} \right) - 
\mu_\infty^2 \, \ln \left( \frac{4 E T}{\mu_\infty^2} \right) \right] \, ,
\eeqar{dEdl_LPM_f} 
where we used $\qm= \sqrt{4 E T}$~\cite{Adil}. Finally, fractional energy loss 
then becomes
\beqar
\frac{\Delta E^{\mathrm{dyn}}_{\mathrm{M}}}{E} =
\frac{C_R \alpha_s}{4\, E}\, \frac{L^2 \mu_\infty^2}{\lambda_\mathrm{dyn}} \,   
    \ln \left( \frac{4 E T}{\mu_E^2} \right)\,,
\eeqar{ELossLPM_M}
\beqar
\frac{\Delta E^{\mathrm{dyn}}_{\mathrm{E}}}{E} \approx 
\frac{\Delta E^{\mathrm{dyn}}_{\mathrm{M}}}{E} 
\frac{\mu_E^2-\mu_\infty^2}{\mu_\infty^2} \,.
\eeqar{ELossLPM_E}
From the above two equations, it directly follows that, in the asymptotic 
limit, the ratio of magnetic and electric contributions to the dynamical QCD 
medium is equal to $\frac{1}{2}$ (i.e. 
$\frac{\mu^2_\infty}{\mu^2_E-\mu^2_\infty}=\frac{1}{2}$). Therefore, for dynamical 
QCD medium (in asymptotic limit), two thirds of the energy loss contribution 
comes from the longitudinal gluon exchange (electric contribution), while one 
third of the contribution comes from the transverse gluon exchange (magnetic 
contribution). 

It was previously shown~\cite{DH_PRL,MD_PRC} that, at asymptotically large 
jet energies, static and dynamical energy losses become equal up to a 
multiplicative constant $\frac{\lambda_\mathrm{dyn}}{\lambda_\mathrm{stat}}$ that 
can be renormalized. One should note that, in static QCD medium, only 
longitudinally polarized gluons give rise to the energy loss. Therefore, we 
have shown here that, while at asymptotically high energies, 
dynamical and static energy losses become numerically equal, their physical 
origin is different: For dynamical medium both longitudinally and transversely 
polarized gluons contribute to the energy loss, while for static medium there 
is only contribution from longitudinally polarized gluons.

%%%%%%%%%%%%%%%%%%%%%%%%%%%%%%%%%%%%%%%%%%%%%%%%%%%%%%%%%%%%%%%%%%%%%%%%
\section{Numerical analysis} 
\label{sec4}
%%%%%%%%%%%%%%%%%%%%%%%%%%%%%%%%%%%%%%%%%%%%%%%%%%%%%%%%%%%%%%%%%%%%%%%%

In this section we first use Eqs.~(\ref{vTL_fin}) and~(\ref{DeltaEMagnElec}) to 
study the importance of magnetic and electric contributions to the energy loss 
in dynamical QCD medium. The analysis is done in the energy range relevant for 
both RHIC and LHC experiments, and for the case of light, charm and bottom 
quarks.

In the numerical analysis we use the following parameters: We consider a 
quark-gluon plasma of temperature $T{\,=\,}225$\,MeV, with $n_f{\,=\,}2.5$ 
effective light quark flavors, and strong interaction strength 
$\alpha_s{\,=\,}0.3$, as representative of average conditions encountered 
in Au+Au collisions at RHIC. For the light quark jets we assume that their 
mass is dominated by the thermal mass $M{\,=\,}\mu_E/\sqrt{6}$, where 
$\mu_E{\,=\,}gT\sqrt{1{+}N_f/6}\approx 0.5$ GeV is the Debye screening 
mass. The charm mass is taken to be $M{\,=\,}1.2$\,GeV, and for 
the bottom mass we use $M{\,=\,}4.75$\,GeV.  To simulate (average) 
conditions in Pb+Pb collisions at the LHC, we use the temperature of 
the medium of $T{\,=\,}400$\,MeV.

%%%%%%%%%%%%%%%%%%%%%%%% Fig. 2 %%%%%%%%%%%%%%%%%%%%%%%%%%%%%%%%%%%%%%%%%%%%
\begin{figure}[ht]
\vspace*{5.5cm} 
\includegraphics{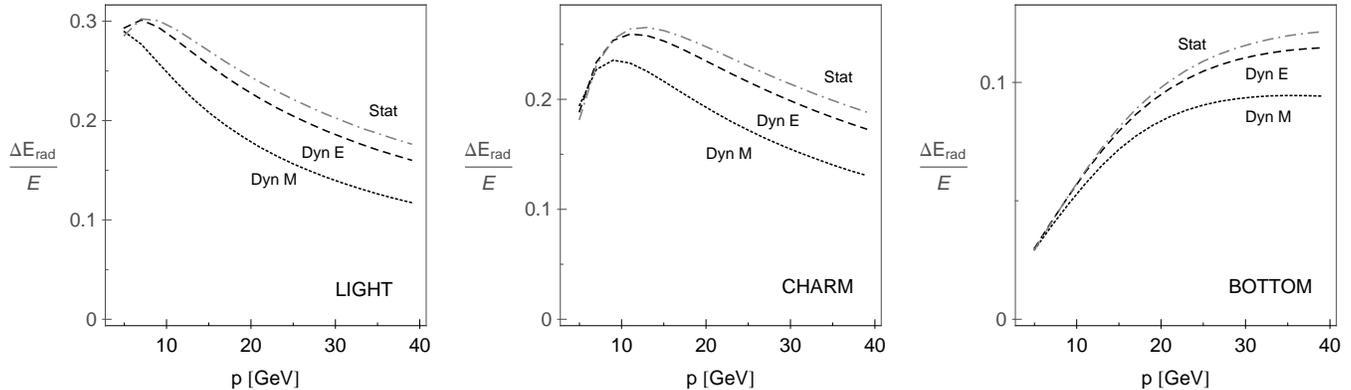}
\caption{Fractional radiative energy loss as a function of momentum for an 
assumed path length $L=5$\,fm and a medium of temperature $T=225$\,MeV (``RHIC 
conditions''). Left, center and right panels correspond respectively 
to light, charm and bottom quarks. Dotted and dashed curves correspond, 
respectively, to magnetic and electric contributions to the energy loss in a 
dynamical QCD medium, while dot-dashed curve corresponds to the energy loss 
in a static QCD medium (see~\cite{DG_Ind}).}
\label{Magn_vs_Elec_RHIC}
\end{figure}
%%%%%%%%%%%%%%%%%%%%%%%%%%%%%%%%%%%%%%%%%%%%%%%%%%%%%%%%%%%%%%%%%%%%%%%%%%%%

\begin{figure}[ht]
\vspace*{5.2cm} 
\includegraphics{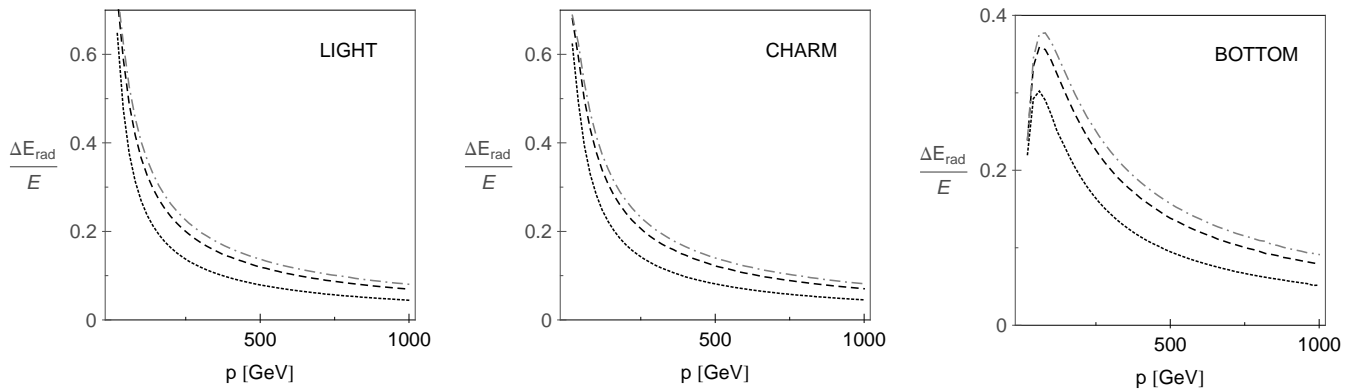}
\caption{Fractional radiative energy loss as a function of momentum for an 
assumed path length $L=5$\,fm and a medium of temperature $T=400$\,MeV (``LHC 
conditions''). Left, center and right panels correspond respectively 
to light, charm and bottom quarks. Dotted and dashed curves correspond, 
respectively, to magnetic and electric contributions to the energy loss in a 
dynamical QCD medium, while dot-dashed curve corresponds to the energy loss 
in a static QCD medium (see~\cite{DG_Ind}).}
\label{Magn_vs_Elec_LHC}
\end{figure}

In Figs.~\ref{Magn_vs_Elec_RHIC} and~\ref{Magn_vs_Elec_LHC} we compare the 
momentum dependence of the magnetic and electric contributions to the radiative 
energy loss in dynamical QCD medium. Figure~\ref{Magn_vs_Elec_RHIC} corresponds 
to RHIC conditions, while Fig.~\ref{Magn_vs_Elec_LHC} corresponds to LHC 
conditions. The figures show that for all three types of quarks, and for both 
RHIC and LHC conditions, {\em i}) both electric and magnetic contributions play 
a significant part in the energy loss, {\em ii}) electric contribution is 
similar (though somewhat smaller) compared to the total energy loss in static 
QCD medium (note that in static QCD medium, only electric contribution to the 
energy loss exists~\cite{DH_Inf}). This leads to the conclusion that the 
increase in the energy loss in dynamical (relative to the static) QCD medium, 
exclusively comes from the magnetic contribution.

%%%%%%%%%%%%%%%%%%%%%%%% Fig. 2 %%%%%%%%%%%%%%%%%%%%%%%%%%%%%%%%%%%%%%%%%%%%
%%%%%%%%%%%%%%%%%%%%%%%%%%%%%%%%%%%%%%%%%%%%%%%%%%%%%%%%%%%%%%%%%%%%%%%%%%%%

%%%%%%%%%%%%%%%%%%%%%%%% Fig. 3 %%%%%%%%%%%%%%%%%%%%%%%%%%%%%%%%%%%%%%%%%%%%
\begin{figure}
\vspace*{5.4cm} 
\includegraphics{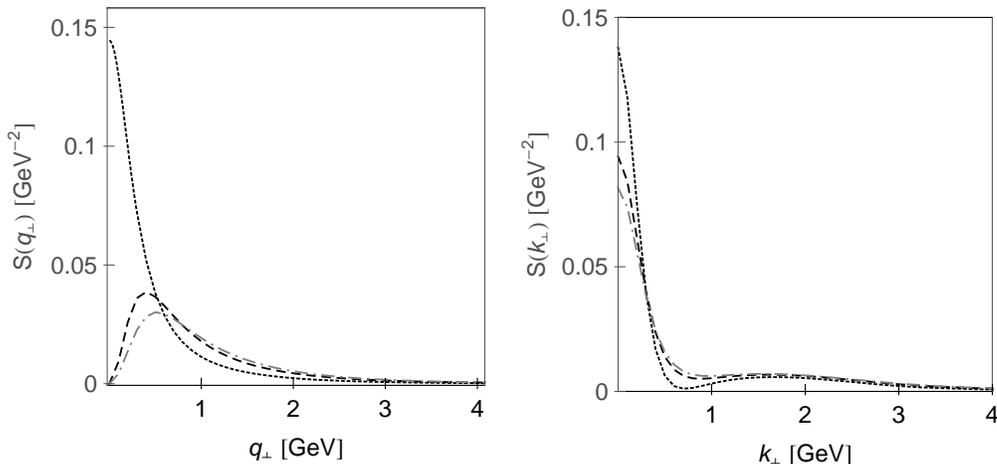}
\caption{Transverse momentum spectrum of the exchanged 
($S(\bq) \equiv \frac{1}{E}\frac{dE}{d^2 \bq}$, left panel) and emitted 
($S(\bk) \equiv \frac{1}{E}\frac{dE}{d^2 \bk}$, right panel) gluons for light 
quarks traveling for $L=5$\,fm through 
a dynamical or static QCD medium. Assumed temperature of the medium is 
$T=225$\,MeV (``RHIC conditions''). Dotted and dashed curves correspond, 
respectively, to magnetic and electric contributions to the transverse 
momentum spectrum in a dynamical QCD medium, while dot-dashed curve 
corresponds to the momentum spectrum in a static QCD medium.
The initial charm quark energy is assumed to be 20 GeV. Note that 
$|\bq|$ and $|\bk|$ are denoted as q$_\perp$ and k$_\perp$ in the figure.}
\label{kqdep}
\end{figure}
%%%%%%%%%%%%%%%%%%%%%%%%%%%%%%%%%%%%%%%%%%%%%%%%%%%%%%%%%%%%%%%%%%%%%%%%%%%%

In Figure~\ref{kqdep} we plot the transverse momentum dependence of the 
exchanged (the left panel) and emitted (the right panel) gluon spectrum. For 
both exchanged and emitted gluon spectrum, we observe similar behavior of the 
electric contribution in dynamical and static QCD medium; this is consistent 
with the similarity between electric contribution to the energy loss in 
dynamic and static QCD medium shown in Figs.~\ref{Magn_vs_Elec_RHIC} 
and~\ref{Magn_vs_Elec_LHC}. On the other hand, we observe a qualitatively 
different momentum dependence for electric and magnetic contributions in the 
exchanged momentum spectrum (the left panel in Fig.~\ref{kqdep}). In 
particular, we see that magnetic contribution to the energy loss comes almost 
exclusively from low momentum spectrum (where $|\bq| \lesssim1$\,GeV). This 
difference between the electric and magnetic contributions does not appear in 
the emitted gluon spectrum (see the right panel in Fig.~\ref{kqdep}), which is 
a consequence of the fact that the difference between electric and magnetic 
contributions comes exclusively from $ v_{T, \, L}(\bq)$ (that does not depend 
on $\bk$ and is a property of the exchanged gluon only).

%%%%%%%%%%%%%%%%%%%%%%%% Fig. 4 %%%%%%%%%%%%%%%%%%%%%%%%%%%%%%%%%%%%%%%%%%%%
\begin{figure}
\vspace*{5.4cm} 
\includegraphics{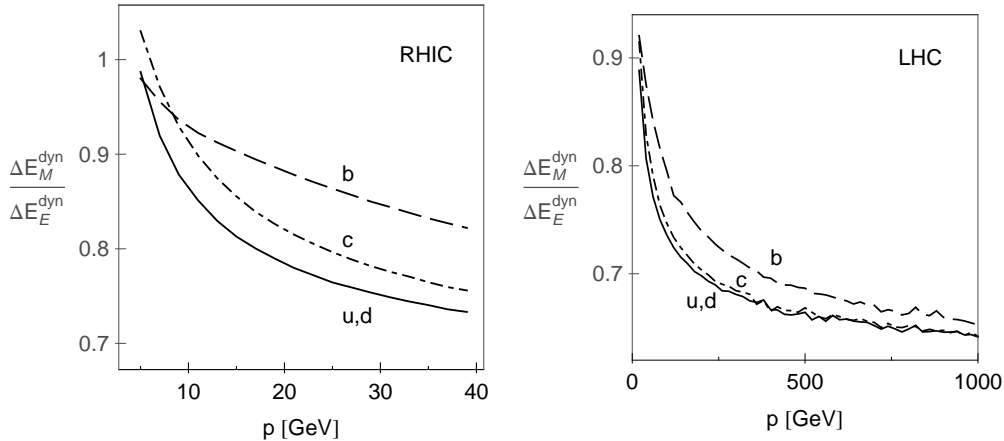}
\caption{Ratio of magnetic and electric contributions to the radiative energy 
loss in a finite size dynamical QCD medium for light, charm and bottom quark 
(full, dot-dashed and dashed curve, respectively). Left (right) panel 
corresponds to the RHIC (LHC) case, with assumed temperature $T=225$\,MeV 
($T=400$\,MeV).}
\label{ratio}
\end{figure}
%%%%%%%%%%%%%%%%%%%%%%%%%%%%%%%%%%%%%%%%%%%%%%%%%%%%%%%%%%%%%%%%%%%%%%%%%%%%

In Fig.~\ref{ratio} we analyze relative importance of magnetic and electric 
contributions for increasingly large values of jet energy. In order to see how 
the asymptotic limit discussed in the previous subsection is approached for 
energies at RHIC and LHC. Right (left) panel of the figure corresponds to the 
RHIC (LHC) case. For both cases, we see that relative importance of magnetic 
contribution decreases with the increase in jet energy, in similar way for all 
three types of quarks. We note that, for asymptotically large jet energies, the 
ratio will reach a limiting value of $1/2$, consistently with the analytical 
result shown in the previous section (data not shown).

\section{Summary}

In this paper we studied the origin of the energy loss increase in dynamical 
QCD medium relative to static QCD medium. While energy loss in static QCD 
medium has only electric contribution from the gluon exchange, we find that, 
in dynamical QCD medium, both electric and magnetic contributions exist and 
are comparable. Since electric contributions in static and dynamical QCD 
medium are approximately equal at RHIC and LHC energies, the boost of energy 
loss results relative to the static QCD medium comes entirely from the 
additional magnetic contribution in the dynamical QCD medium. Furthermore, 
while energy loss in dynamical QCD medium approaches the static results 
for asymptotically high energies, the physical origin of the energy loss is 
different for dynamical and static case in this limit: In the static case, 
the entire energy loss comes from the electric contribution, while for dynamical 
medium one third of the energy loss comes from magnetic contribution.

The calculation presented here is based on HTL perturbative QCD, which requires 
zero magnetic mass. On the other hand, different non-perturbative 
approaches~\cite{Maezawa,Nakamura,Hart,Bak} suggest a non-zero magnetic mass for 
RHIC and LHC case. Since this paper established significance of the magnetic 
contribution in the radiative energy loss, the work presented here therefore 
opens the following question: Is it possible to consistently include finite 
magnetic mass in the dynamical energy loss calculations, and how this inclusion 
would modify energy loss result? Addressing this question is our next important 
goal.
\begin{acknowledgments}
This work is supported by Marie Curie International Reintegration Grant within the 
$7^{th}$ European Community Framework Programme (PIRG08-GA-2010-276913) and by the 
Ministry of Science and Technological Development of the Republic of Serbia,
under projects No. ON171004 and ON173052.
\end{acknowledgments}

%%%%%%%%%%%%%%%%%%% References %%%%%%%%%%%%%%%%%%%%%%%%%%%%%%%%%%%%%%%%%%%%%

\end{document}